\documentclass[a4paper,11pt]{article}
\usepackage{pos}
\pdfoutput=1

\usepackage{subcaption} 

\title{Searches for lepton flavour and lepton number violation in $K^{+}$ decays at NA62}
 \ShortTitle{Searches for LFV/LNV in $K^{+}$ decays at NA62}

\author*[\dagger]{Joel Swallow}

\affiliation[]{The University of Birmingham,\\
  School of Physics and Astronomy, Edgbaston B15 2TT, UK}

\emailAdd{joel.christopher.swallow@cern.ch}

\notes{\note{
for the NA62 Collaboration:

R.~Aliberti, F.~Ambrosino, R.~Ammendola, B.~Angelucci, A.~Antonelli, G.~Anzivino, R.~Arcidiacono, T.~Bache, A.~Baeva, D.~Baigarashev
% ...

}}

\abstract{A search for the lepton number violating $K^{+}\rightarrow\pi^{-}\mu^{+}e^{+}$ and lepton flavour violating  $K^{+}\rightarrow\pi^{+}\mu^{-}e^{+}$ decays is reported using NA62 data collected in 2017 and 2018. No significant excess is observed above the background expectation and therefore upper limits are established on the branching ratios at $90\%$ confidence level: $\mathcal{B}(K^{+}\rightarrow\pi^{-}\mu^{+}e^{+})<4.2\times10^{-11}$ and $\mathcal{B}(K^{+}\rightarrow\pi^{+}\mu^{-}e^{+})<6.6\times10^{-11}$. These results improve over the previous limits by factors of $12$ and $8$ respectively.}

\FullConference{%
  40th International Conference on High Energy physics - ICHEP2020\\
  July 28 - August 6, 2020\\
  Prague, Czech Republic (virtual meeting)
}

\begin{document}
\maketitle

\section{Introduction}
In the standard model (SM) the lepton number $L$ and lepton flavour numbers $L_{\ell},\,\ell\in[e,\mu,\tau]$ are conserved due to apparent approximate global symmetries. However, this is an emergent property of the SM not required during its construction and evidence for violation of lepton flavour numbers (LFV) has been demonstrated by the observation of neutrino oscillations. No observation of LFV for charged leptons or any lepton number violating (LNV) process has been observed, and either would be a clear indication of physics beyond the SM (BSM). 

A number of postulated BSM scenarios include mechanisms resulting in LFV and LNV. Seesaw models constructed to explain the mass of neutrinos, orders of magnitude lower than other fundamental fermions, introduce Majorana neutrino mass terms and exchange of Majorana neutrinos can facilitate LNV ($\Delta L=2$) processes~\cite{LittenbergShrock00}~\cite{Atre09}. In other BSM scenarios the existence of leptoquarks is postulated which can couple to both leptons and quarks, exchange of these particles can then allow LFV processes ($\Delta L_{e}=1$, $\Delta L_{\mu}=1$)~\cite{CahnHarari80}~\cite{MandalPich19}. 

The NA62 experiment at CERN has collected the largest sample of $K^{+}$ decays to date and this data has been analysed to search for LNV and LFV processes.  
NA62 has set new upper limits on LNV $K^{+}\rightarrow\pi^{-}\ell^{+}\ell^{+}$ decays, of $2.2\times10^{-10}$ and $4.2\times10^{-11}$ (at $90\%$ confidence level) in the cases of $\ell=e$ and $\ell=\mu$ respectively using a sub-set of 2017 data, representing $30\%$ of the data collected to date (Run 1)~\cite{NA62_Kpill_LNV}. 
In this proceedings contribution the search for $K^{+}\rightarrow\pi^{\mp}\mu^{\pm}e^{+}$ decays using the full Run 1 data-set is reported. 

\section{The NA62 beamline and detector}
\label{sec:Detector}
\begin{figure}
    \centering
    \vspace{-5pt}
    \includegraphics[width=1.0\textwidth]{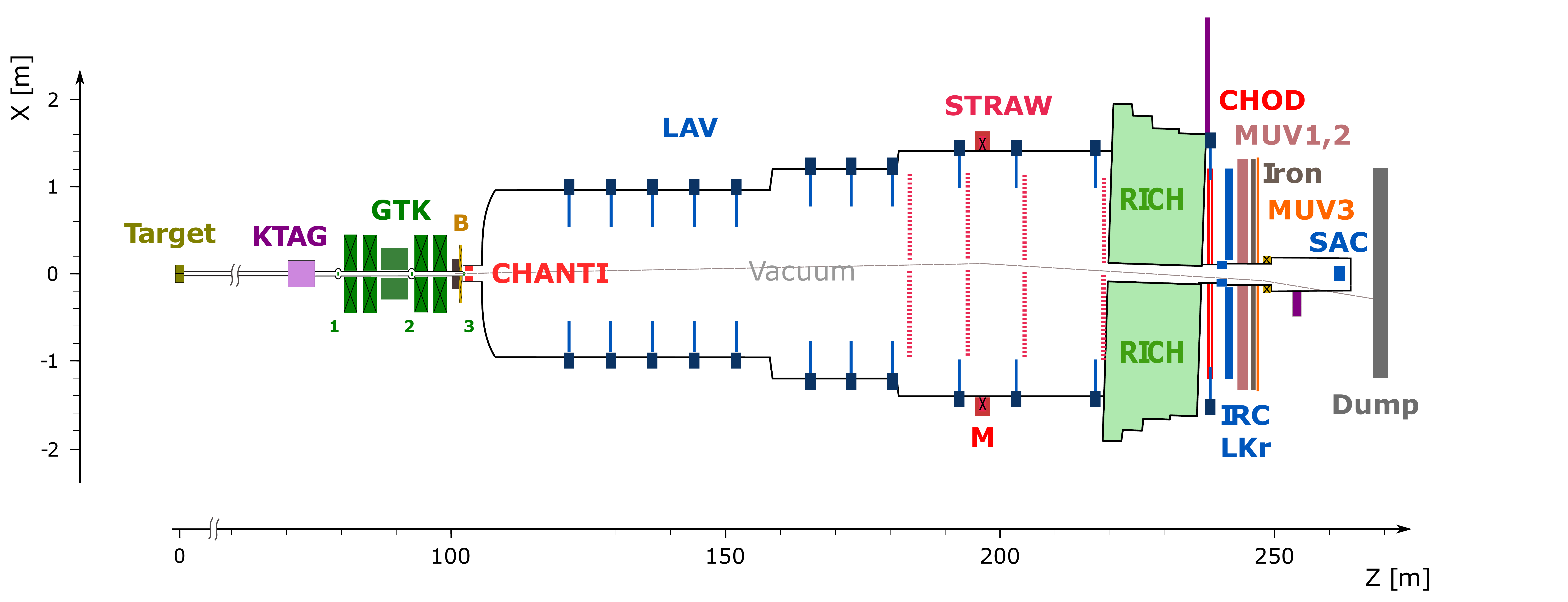}
    \caption{Schematic top view of the NA62 beamline and detector.}
    \label{fig:Detector}
\end{figure}

The NA62 experiment, beamline and detector are described in detail in~\cite{NA62DetectorPaper} and a schematic diagram is shown in figure~\ref{fig:Detector}.
A high intensity $400\,\text{GeV}/c$ proton beam is delivered to the NA62 experiment by the CERN SPS in spills of $3\,s$ effective duration containing $\sim1.8\times10^{13}$ protons. The beam impinges onto a beryllium target producing a secondary hadron beam composed of $\pi^{+}\,(70\%)$ protons $(23\%)$ and $K^{+}\,(6\%)$, with mean momentum of $75\,\text{GeV}/c$ and a $1\%$ rms spread. The KTAG, a differential Cherenkov counter, positively tags $K^{+}$ with a time resolution better than $70\,\text{ps}$. Kaon decays in a $75\,\text{m}$ fiducial volume (FV), contained in a large vacuum tank, are studied. A magnetic spectrometer (STRAW), formed of four straw tracking chambers and a dipole magnet (M), is used to reconstruct the momenta of charged particles produced in $K^{+}$ decays in the FV. The CHOD, constructed from two sets of scintillator hodoscope planes, provides time measurements for charged tracks with $200\,\text{ps}$ precision. 
For three-charged-track final states a vertex can be reconstructed in the FV, and associated to a $K^{+}$ decay if the total final state momentum is consistent with the (measured) average beam momentum, and is coincident in time with a KTAG $K^{+}$ tag signal. This means studies of three-charged-track decays do not necessarily require measurements of the beam $K^{+}$ momentum, provided by the gigatracker (GTK) and crucial to the study of the $K^{+}\rightarrow\pi^{+}\nu\bar{\nu}$ decay~\cite{PNN17}.
Information for particle identification (PID), to distinguish $\pi^{\pm},\,\mu^{\pm}$ and $e^{\pm}$, is provided by a quasi-homogeneous liquid krypton calorimeter (LKr) and muon detector (MUV3). A photon veto system is formed from twelve large angle veto (LAV) stations, the LKr and small angle vetos (IRC and SAC). The ring imaging Chernkov detector (RICH), CHOD, LKr and MUV3 provide signals used for triggering. 

\section{Analysis Strategy}
A search for $K^{+}\rightarrow\pi^{\mp}\mu^{\pm}e^{+}$ decays has been performed using 2017 and 2018 NA62 data, equivalent to the full Run 1 data-set for these analyses. Potential signal events may be identified if the three-charged-track $\pi^{\mp}\mu^{\pm}e^{+}$ candidate final state has an invariant mass, $M_{\pi\mu e}$, consistent with the $K^{+}$ mass. A blind analysis strategy is adopted and the $M_{\pi\mu e}$ range $478$--$510\,\text{MeV}/c^{2}$ was masked until the analysis was finalised. The signal region, $490$--$498\,\text{MeV}/c$, was chosen, centred on the $K^{+}$ mass and with width approximately equal to $6$ times the invariant mass resolution. Two independent analyses were performed with results cross-validated. 
The most common three-track $K^{+}$ decay, $K^{+}\rightarrow\pi^{+}\pi^{+}\pi^{-}$, is used for normalisation. The core of the selection for normalisation and signal events is identical, requiring a three-track final state with vertex in the FV and consistent with a $K^{+}$ decay (see section~\ref{sec:Detector}) and vetoing photons detected by the LAVs. Following the common selection criteria normalisation events are identified with a requirement of invariant mass under the $3\pi$ hypothesis consistent with the $K^{+}$ mass, and PID conditions are applied to identify signal candidates with $\pi^{\mp}\mu^{\pm}e^{+}$ final states. For the $K^{+}\rightarrow\pi^{-}\mu^{+}e^{+}$ search it is additionally required that the invariant mass of the identified $\pi^{-}e^{+}$ pair under the $e^{-}e^{+}$ hypothesis $M_{\pi e}>140\,\text{MeV}/c^{2}$, rejecting backgrounds from decay chains involving $\pi^{0}\rightarrow e^{+}e^{-}\gamma$ (for example $K^{+}\rightarrow\pi^{+}[e^{+}e^{-}\gamma]_{\pi^{0}}$) and misidentification. 

Data from three triggers were used for these searches: the minimum bias three-track `Multi-Track' trigger, and specialised `Multi-Track $\mu$' and `Multi-Track $e$' triggers. The Multi-Track $\mu$ trigger saves events with three charged tracks, $10\,\text{GeV}$ of energy deposition in the LKr and at least one time-coincident signal in the MUV3. The Multi-Track $e$ trigger saves three-charged-track events with $20\,\text{GeV}$ energy deposition in the LKr. These triggers are run concurrently with the main `PNN' trigger, used for the study of the $K^{+}\rightarrow\pi^{+}\nu\bar{\nu}$ decay~\cite{PNN17}, but downscaled by factors of approximately $100,8$ and $8$ respectively. 

All three triggers are used to collect candidate signal $K^{+}\rightarrow\pi^{\mp}\mu^{\pm}e^{+}$ events while normalisation $K^{+}\rightarrow\pi^{+}\pi^{+}\pi^{-}$ decays are collected with only the Multi-Track trigger.
The effective number of $K^{+}$ decays in the FV that can be used for these searches, measured using the number of selected normalisation $K^{+}\rightarrow\pi^{+}\pi^{+}\pi^{-}$ decays, the branching ratio of this decay and the downscaling factors and inefficiencies of the triggers, is $(1.32\pm0.01)\times10^{12}$.

\section{Background Studies}
\begin{figure}
    \centering
    \includegraphics[width=0.67\textwidth]{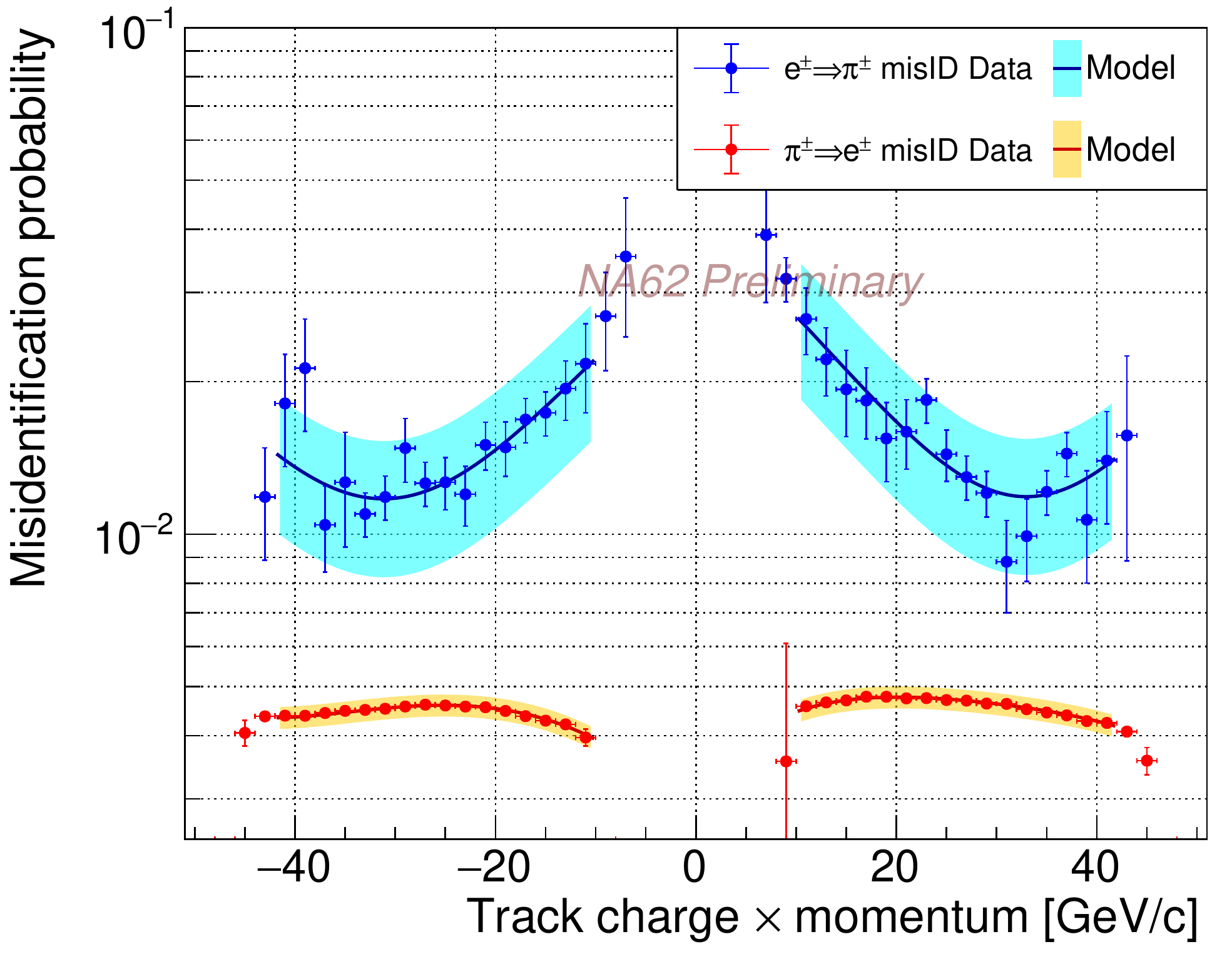}
    \caption{Misidentification probabilities between $\pi^{\pm}$ and $e^{\pm}$ as a function of track charge times momentum.}
    \label{fig:MisID}
\end{figure}

There are two primary processes which can lead to background mechanisms: misidentification (misID) and decays in flight (DIF). 
For example, in both $K^{+}\rightarrow\pi^{\mp}\mu^{\pm}e^{+}$ searches the most significant background at low $M_{\pi\mu e}$ arises from $K^{+}\rightarrow\pi^{+}\pi^{+}\pi^{-}$ decays followed by one $\pi^{\pm}\rightarrow\mu^{\pm}\nu_{\mu}$ DIF and one misID of a $\pi^{+}$ as an $e^{+}$. Simulations reproduce the effect of the DIF processes observed in data, however for misID data-driven models must be used. The probability of misID of a $\pi^{+}$ as an $e^{+}$ is measured using a control sample of $K^{+}\rightarrow\pi^{+}\pi^{+}\pi^{-}$ decays collected with the Multi-Track trigger. A model is constructed to describe the misID probability as a function of track momentum as shown in figure~\ref{fig:MisID}.

Similarly, $K^{+}\rightarrow\pi^{+}e^{+}e^{-}$ decays with an $e^{-}$ misidentified as a $\pi^{-}$ (and DIF or misID of the $\pi^{+}$ to produce a $\mu^{+}$) can lead to a background in the signal region for the $K^{+}\rightarrow\pi^{-}\mu^{+}e^{+}$ search. In this case a control sample of $e^{\pm}$ from $K^{+}\rightarrow\pi^{+}[e^{+}e^{-}\gamma]_{\pi^{0}}$ decays is used to measure the misID probability of $e^{\pm}$ as a $\pi^{\pm}$, as shown in figure~\ref{fig:MisID}. 

Background expectations for the $K^{+}\rightarrow\pi^{\mp}\mu^{\pm}e^{+}$ searches are, in general, evaluated using simulations with significant data-driven corrections, such as the data-driven misID models discussed above. Results are shown in tables~\ref{tab:BkgPred_piM} and~\ref{tab:BkgPred_muM} for the $K^{+}\rightarrow\pi^{-}\mu^{+}e^{+}$ and $K^{+}\rightarrow\pi^{+}\mu^{-}e^{+}$ searches respectively.
Before unblinding the expected number of background events at high $M_{\pi\mu e}$, above the blinded region, is found to be compatible with the observed number of events. To validate the background predictions for the signal region the side-bands within the blinded region but outside the signal region were used as control regions, again good agreement is obtained between the expected and observed number of events. The predicted numbers of background events in the signal regions of the $K^{+}\rightarrow\pi^{-}\mu^{+}e^{+}$ and $K^{+}\rightarrow\pi^{+}\mu^{-}e^{+}$ searches are $1.06\pm0.20$ and $0.92\pm0.34$ respectively.

\begin{table}
    \caption{Number of expected background and observed events for $K^{+}\rightarrow\pi^{-}\mu^{+}e^{+}$ search.}
    \centerline{
    \resizebox{1.05\textwidth}{!}{
        \begin{tabular}{|c|c|c|c|c|}
            \hline
            Region
            & \multicolumn{1}{c|}{High mass Region} & \multicolumn{2}{c|}{Control Regions (Sidebands)} & Signal Region \\
            \multicolumn{1}{|c|}{$M_{\pi\mu e}$ range $[\text{MeV}/c^{2}]$} & $>510$ & $478$--$490$ & $498$--$510$ & $490$--$498$ \\
	    \hline
            Total background expected
                & $5.50\pm0.53$ 
                & $1.68\pm0.20$ & $1.66\pm0.26$
                & $1.06\pm0.20$ \\
            \hline
            Data & $8$ & $2$ & $4$ & $0$ \\
            \hline
        \end{tabular}
    }
    }
    \label{tab:BkgPred_piM}
\end{table}

\begin{table}
    \caption{Number of expected background and observed events for $K^{+}\rightarrow\pi^{+}\mu^{-}e^{+}$ search.}
    \centerline{
    \resizebox{1.05\textwidth}{!}{
        \begin{tabular}{|c|c|c|c|c|}
            \hline
            Region
            & \multicolumn{1}{c|}{High mass Region} & \multicolumn{2}{c|}{Control Regions (Sidebands)} & Signal Region \\
            \multicolumn{1}{|c|}{$M_{\pi\mu e}$ range $[\text{MeV}/c^{2}]$} & $>510$ & $478$--$490$ & $498$--$510$ & $490$--$498$ \\
	    \hline
            Total background expected
                & $1.95\pm0.48$ 
                & $3.41\pm0.54$ & $1.27\pm0.40$
                & $0.92\pm0.34$ \\
            \hline
            Data & $4$ & $2$ & $0$ & $2$ \\
            \hline
        \end{tabular}
    }
    }
    \label{tab:BkgPred_muM}
\end{table}

\section{Results}
\begin{figure}
    \centering
    \begin{subfigure}[b]{0.49\textwidth}
        \includegraphics[width=1.0\textwidth]{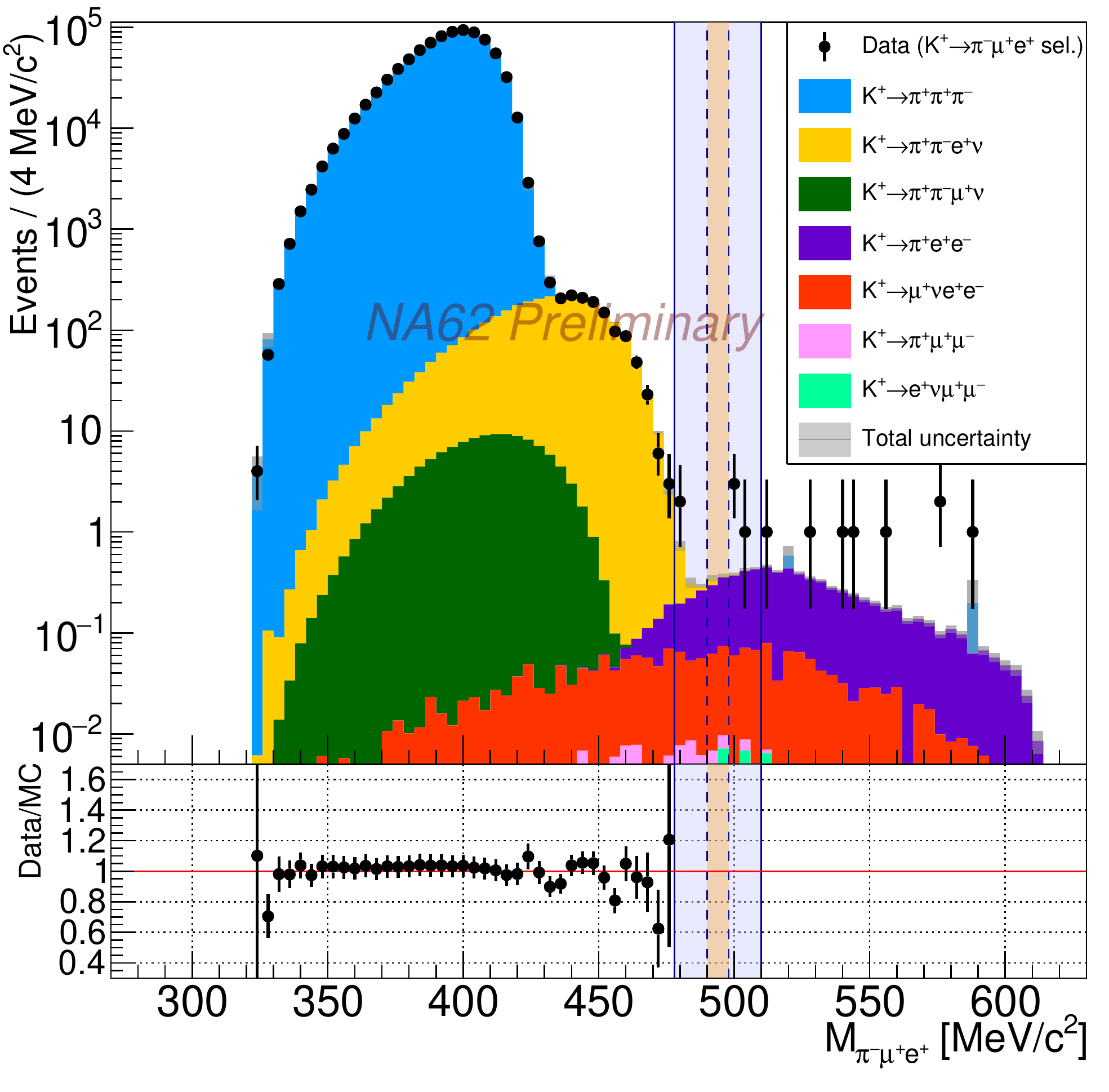}
        \caption*{} 
        \label{fig:Kpimue_piM} 
    \end{subfigure}
    \begin{subfigure}[b]{0.49\textwidth}
        \includegraphics[width=1.0\textwidth]{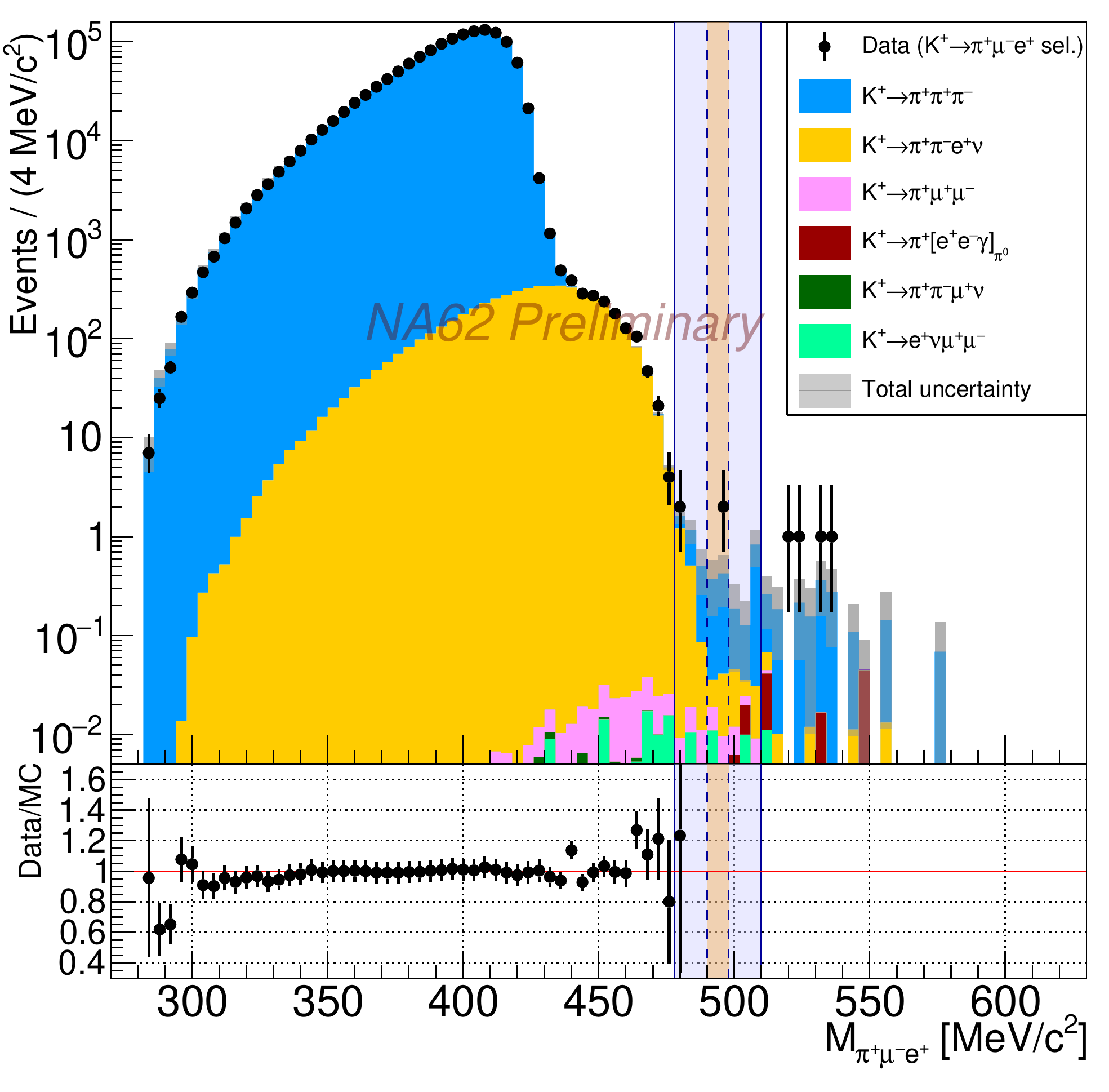} 
        \caption*{}
        \label{fig:Kpimue_muM}
    \end{subfigure}
    \vspace{-15pt}
    \caption{
    Reconstructed mass spectra for selected events in searches for $K^{+}\rightarrow\pi^{-}\mu^{+}e^{+}$ (left) and $K^{+}\rightarrow\pi^{+}\mu^{-}e^{+}$ (right) for data and simulated samples.
    } 
    \label{fig:Mpimue}
\end{figure}

After unmasking the signal region $0$ and $2$ events are found for the $K^{+}\rightarrow\pi^{-}\mu^{+}e^{+}$ and $K^{+}\rightarrow\pi^{+}\mu^{-}e^{+}$ searches respectively, with full invariant mass spectra shown in figure~\ref{fig:Mpimue}. This is compatible with the background expectations and therefore upper limits are established on the branching ratios, using the CLs method~\cite{Read02}, at:
\begin{align}
    \mathcal{B}(K^{+}\rightarrow\pi^{-}\mu^{+}e^{+})<4.2\times10^{-11}\,@\,90\%\,CL\,, \\
    \mathcal{B}(K^{+}\rightarrow\pi^{+}\mu^{-}e^{+})<6.6\times10^{-11}\,@\,90\%\,CL\,.
\end{align}
These represent an improvement on the previous limits, set by the BNL E865 experiment~\cite{Appel00}, by factors of $12$ and $8$ respectively.  

\section{Conclusion and Outlook}
Searches for LNV/LFV decays $K^{+}\rightarrow\pi^{\mp}\mu^{\pm}e^{+}$ have been performed using 2017 and 2018 (Run 1) NA62 data. New upper limits on the branching ratios are established, improving over previous constraints by approximately an order of magnitude. 
In general searches for LNV/LFV $K^{+}$ decays at NA62 are not limited by background and with data-taking resuming in 2021 NA62 has strong prospects to improve sensitivities for these searches. 

\bibliographystyle{JHEP}
\bibliography{Bibliography.bib}

\end{document}